\documentclass[aps,prl,floats,showpacs,twocolumn]{revtex4}
\usepackage{graphicx,color}

\begin{document}
\title{Flexural phonons in free-standing graphene}
\author{Eros Mariani}
\author{Felix von Oppen}
\affiliation{Institut f\"ur Theoretische Physik, Freie Universit\"at Berlin, Arnimallee 14, 14195 Berlin, Germany}
\date{\today}
\begin{abstract}
Rotation and reflection symmetries impose that out-of-plane (flexural) phonons of free-standing graphene membranes have a quadratic dispersion at long wavelength and can be excited by charge carriers in pairs only. As a result, we find that flexural phonons dominate the phonon contribution to the resistivity $\rho$ below a crossover temperature $T_x$ where we obtain an anomalous temperature dependence $\rho\propto T^{5/2}_{}\ln T$. The logarithmic factor arises from renormalizations of the flexural phonon dispersion due to coupling between bending and stretching degrees of freedom of the membrane.
\end{abstract}
 \maketitle

{\em Introduction}.---The experimental realization of monolayers of graphite, termed graphene, has opened new horizons in the physics of two-dimensional electron systems (2DES) \cite{Geim,Kim}. Unlike conventional 2DES, the low-energy electronic bandstructure of graphene \cite{Wallace,Gonzales,Dresselhaus} is described by a massless Dirac equation with velocity $v$. The $4\times 4$ matrix structure of the Dirac equation reflects the two sublattices of the graphene honeycomb lattice in combination with a valley degeneracy due to the presence of two Dirac cones within the Brillouin zone. Pioneering experiments on this novel two-dimensional electron system have shown that the Dirac nature of carriers induces an anomalous integer quantum Hall effect as well as a finite conductivity at vanishing carrier density \cite{Geim,Kim}.

Recently, it has become possible to experiment on free-standing graphene sheets \cite{GeimRipples,McEuen} which provide a realization of a two-dimensional (2d) solid. Studies of the stability of 2d solids against thermal fluctuations date back to early work by Peierls and Landau \cite{Peierls, Landau} who pointed out the absence of true long-range translational order. Much later, it was understood \cite{Kosterlitz} that nevertheless, quasi-long-range translational order can persist up to a finite-temperature Kosterlitz-Thouless transition. While these works focus on in-plane distortions of the lattice, free-standing membranes also support out-of-plane distortions. It is believed that there is a low-temperature flat phase even in the presence of out-of-plane distortions \cite{Nelson}, which gives way to a crumpled phase at high temperatures \cite{NelsonCrumpling}. 

Within the low-temperature flat phase, long-wavelength elastic distortions, both in-plane and out-of-plane, can be described by the appropriate elastic Lagrangian density \cite{Nelson,LandauBook},
\begin{equation}
    {\cal L} = \frac{\rho_0}{2} (\dot{\bf u}^2+\dot h^2) -\frac{1}{2}\kappa (\nabla^2 h)^2 - \mu
    u^2_{ij}-\frac{1}{2} \lambda u^2_{kk} 
\label{elastic}
\end{equation}
in terms of the mass density $\rho_0$, the out-of-plane distortions $h({\bf r})$, and the strain tensor $u_{ij}=\frac{1}{2}[\partial_i u_j + \partial_j u_i + (\partial_i h) (\partial_j h)]$. Here, ${\bf u}({\bf r})$ denotes the in-plane distortions. The elastic constants $\lambda$ and $\mu$ characterize the in-plane rigidity of the lattice, $\kappa$ the bending rigidity. Both the absence of a $(\nabla h)^2$ term in the elastic Lagrangian and the appearance of the out-of-plane distortions $h({\bf r})$ in the strain tensor are direct consequences of the rotational symmetry of the membrane in the embedding space. 

The elastic Lagrangian in Eq.\ (\ref{elastic}) encapsulates a distinct difference between in-plane and out-of-plane (flexural) phonons. Indeed, to quadratic order in the displacement fields $h({\bf r})$ and ${\bf u}({\bf r})$, both longitudinal and transverse in-plane phonons have a {\em linear} dispersion $\omega^{(l)}_{\mathbf{q}}=v^{(l)}q$ and $\omega^{(t)}_{\mathbf{q}}=v^{(t)}q$ with group velocities $v^{(l)}=\left[\left(2\mu +\lambda\right)/\rho_0\right]^{1/2}$ and $v^{(t)}=\left[\mu/\rho_0\right]^{1/2}$. In contrast, flexural phonons obey a {\em quadratic} dispersion $\omega_{\mathbf{q}}^{(h)} = \alpha q^2$ with $\alpha =\left[\kappa/\rho_0\right]^{1/2}$, which is a consequence of
rotational symmetry. In-plane and flexural phonons also differ in their coupling to the charge carriers of graphene membranes. While the coupling is conventional for in-plane phonons, the reflection symmetry $h\to -h$ demands that out-of-plane displacements enter only {\em quadratically} into the Dirac Hamiltonian. Consequently, charge carriers can excite flexural phonons only {\em in pairs}. 

Due to these differences, we find that flexural phonons dominate the phonon contribution to the resistivity of free-standing graphene membranes below a crossover temperature $T_x$. Indeed, the transport scattering rate of Dirac fermions {\em diverges} logarithmically for a strictly quadratic dispersion of flexural phonons. This divergence is cut off by the coupling between bending and stretching degrees of freedom of the membrane, as captured by the elastic Lagrangian Eq.\ (\ref{elastic}). At finite temperature, this coupling renormalizes the bending rigidity of the membrane, inducing a stiffening of the flexural-mode dispersion at long wavelengths. Including this physics within a simple one-loop RG, we find that the contribution of flexural phonons to the resistivity of graphene membranes scales as $T^{5/2}_{}\ln T$. 

{\em Graphene}.---The bandstructure of graphene is well approximated by the tight-binding Hamiltonian  
\begin{equation}
\label{H}
H=-t \sum_{\langle ij\rangle} [c^{\dagger}_{i}c_{j}+c^{\dagger}_{j}c_{i}]
\end{equation}
on a honeycomb lattice. Here, $t$ is the hopping matrix element, $c_{i}$ annihilates an electron on lattice site $i$, and only nearest-neighbor hopping has been included. The 2d hexagonal lattice consists of two identical sublattices A and B, and thus two sites per unit cell. We denote the vectors connecting a B site with the neighboring A sites as $\mathbf{e}^{}_{1}=a\, (-1,0)$, $\mathbf{e}^{}_{2}=a\, (1/2,\sqrt{3}/2)$ and $\mathbf{e}^{}_{3}=a\, (1/2,-\sqrt{3}/2)$ where $a$ is the bond length. The band structure of the Hamiltonian in Eq.\ 
(\ref{H}) has zero energy (corresponding to the Fermi energy at half filling) at two inequivalent points in the Brillouin zone, which we choose to be at $\mathbf{k}^{}_{\pm}=\pm\mathbf{k}_{D}^{}$, with $\mathbf{k}_{D}^{}=2\pi/(3\sqrt{3}a)\, (\sqrt{3},1)$. In the vicinity of these Dirac points, the spectrum is described by the $4\times 4$ Dirac Hamiltonian
\begin{equation}
\label{HDirac}
H=\hbar v\,\mathbf{\Sigma}\cdot\mathbf{k}
\end{equation}
with velocity $v=3ta/2$. The 2d wavenumber $\mathbf{k}$ is measured from the Dirac point. The Hamiltonian in Eq.\ (\ref{HDirac}) acts on four-component spinors $(u_{A,\mathbf{k}}^{+},u_{B,\mathbf{k}}^{+},u_{B,\mathbf{k}}^{-},u_{A,\mathbf{k}}^{-})$ of Bloch amplitudes in the space spanned by the sublattices ($A/B$) and Dirac points ($+/-$). The matrices $\Sigma_{x,y}^{}=\Pi_{z}^{}\otimes\mathbf{\sigma}_{x,y}^{}$ denote components of a vector $\mathbf{\Sigma}$. ($\Pi_{i}^{}$ and $\sigma_{j}^{}$ are Pauli matrices acting in the spaces of the Dirac points and the sublattices, respectively). It is also useful to introduce a corresponding vector ${\bf \Lambda}$ by 
$\Lambda_{x,y}^{}=\Pi_{x,y}^{}\otimes\sigma_{z}^{}$ and $\Lambda_{z}^{}=\Pi_{z}^{}\otimes\sigma_{0}^{}$
\cite{FalkoWL}.

\begin{figure}[t]
	\centering
		\includegraphics[width=0.8\columnwidth]{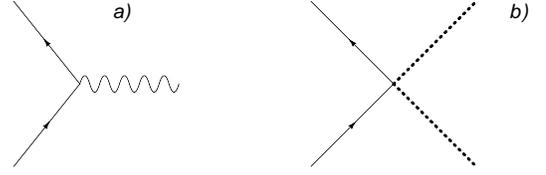}
			\caption{Electron-phonon vertices. a) Coupling of electrons to in-plane phonons. b) Coupling of electrons to flexural phonons. Straight lines correspond to electrons, wavy (dashed) lines to in-plane (flexural) phonons.
			\label{Vertices}}
\end{figure}

{\em Electron-phonon coupling}.---The dominant electron-phonon coupling arises from distortion-induced modifications of the bond lengths and hence the hopping amplitude \cite{Mahan,Manes,Arpes}. The electron-phonon coupling can then be described in terms of a fictitious gauge field ${\bf A}({\bf r})$ entering into the Dirac Hamiltonian, $H=v{\bf \Sigma}\cdot ({\bf p}+e{\bf A}({\bf r},t))$, where $\mathbf{p}$ denotes the momentum. Expressed in terms of the strain tensor $u_{ij}$, the gauge field ${\bf A}({\bf r})$ takes the form \cite{Katsnelson}
\begin{eqnarray}
  e{\bf A}({\bf r},t)&=& \Pi_z \otimes {\bf 1}\, \frac{\hbar}{t}\frac{\partial t}{\partial a}\left[\begin{array}{c} u_{xy} \\
  \frac{1}{2}(u_{xx}-u_{yy})
  \end{array}\right] \; .
  \label{fictitious}
\end{eqnarray}
The combination of components of the strain tensor entering into ${\bf A}({\bf r})$ reflects the symmetry of the underlying honeycomb lattice. The factor of $\Pi_z$ implies that the associated fictitious magnetic field points in opposite directions at the two Dirac points, as required by time-reversal symmetry. From the definition of the strain tensor, we directly conclude that electrons couple linearly to the in-plane phonons and quadratically to flexural ones. The corresponding electron-phonon vertices are depicted in Fig. \ref{Vertices}. 

At sufficiently low temperatures, only long-wavelength phonons contribute to the resistivity and scattering between different Dirac cones can be neglected. Thus, we can restrict attention to the vicinity of, say, the Dirac point $\mathbf{k}_{+}^{}$ where the Hamiltonian reduces to a $2\times 2$ Hamiltonian in the $A-B$ sublattice space. 

The electron-phonon coupling $H_{\mathrm{ep}}^{}$ corresponding to Eq.\ (\ref{fictitious}) can be expressed in second quantization after expanding the in-plane and out-of-plane distortions into Fourier series as $\mathbf{u}(\mathbf{r}) = \sum_{\mathbf{q}}^{}\mathbf{u}_{\mathbf{q}}^{}\, e^{i\mathbf{q}\cdot\mathbf{r}}_{}$ and $h(\mathbf{r}) = \sum_{\mathbf{q}}^{}h_{\mathbf{q}}^{}\, e^{i\mathbf{q}\cdot\mathbf{r}}_{}$, decomposing $\mathbf{u}_{\mathbf{q}}^{}$ into longitudinal and transverse components,
\begin{equation}
\label{ulongtrans}
\mathbf{u}_{\mathbf{q}}^{}=u_{\mathbf{q}}^{(l)}\hat{\mathbf{q}}+u_{\mathbf{q}}^{(t)}\hat{\mathbf{z}}\times\hat{\mathbf{q}}\; ,
\end{equation}
and quantizing the amplitude of the distortions as
\begin{equation}
u_{\mathbf{q}}^{(\nu )} = \sqrt{\frac{\hbar}{2M\omega_{\mathbf{q}}^{(\nu )}}}\,\left(a^{(\nu )}_{\mathbf{q}}+a^{(\nu )\dagger}_{-\mathbf{q}}\right)\;  .
\end{equation}
Here $M$ is the atomic mass and $a^{(\nu )}_{\mathbf{q}}$ the annihilation operator of a phonon of type $\nu$ ($\nu=l,t,h$) with wavenumber $\mathbf{q}$. This yields
\begin{eqnarray}
H_{\mathrm{ep}}^{}=\sum_{\mathbf{k},\mathbf{k}^{\prime},\mathbf{q}}^{}\sum_{\nu =l,t}^{}\, V^{(\nu )}_{\mathbf{q}}c^{\dagger}_{B,\mathbf{k}}c^{}_{A,\mathbf{k}^{\prime}}\left(a^{(\nu )}_{\mathbf{q}}+a^{(\nu )\dagger}_{-\mathbf{q}}\right)\delta_{\mathbf{k}^{\prime},\mathbf{k}^{}-\mathbf{q}^{}}^{} && \nonumber \\
+\sum_{\mathbf{k},\mathbf{k}^{\prime},\mathbf{q}^{}_{},\mathbf{q}^{\prime}}\, V^{(h)}_{\mathbf{q},\mathbf{q}^{\prime}}c^{\dagger}_{B,\mathbf{k}}c^{}_{A,\mathbf{k}^{\prime}}\left( a^{(h)}_{\mathbf{q}}+a^{(h)\dagger}_{-\mathbf{q}}\right)\quad &&\\
\quad\quad\times\left( a^{(h)}_{\mathbf{q}^{\prime}}+a^{(h)\dagger}_{-\mathbf{q}^{\prime}}\right)\delta_{\mathbf{k}^{\prime},\mathbf{k}^{}-\mathbf{q}^{}-\mathbf{q}^{\prime}}^{} +h.c. \quad &&\nonumber
\label{Helph}
\end{eqnarray}
in terms of the coupling terms
\begin{eqnarray}
V^{(l)}_{\mathbf{q}}=&&-\epsilon \, q\, e^{-2i\phi}_{}\sqrt{\frac{\hbar}{2M\omega_{\mathbf{q}}^{(l)}}}\nonumber \\
V^{(t)}_{\mathbf{q}}=&&-i\epsilon \, q\, e^{-2i\phi}_{}\sqrt{\frac{\hbar}{2M\omega_{\mathbf{q}}^{(t)}}}\\
V^{(h)}_{\mathbf{q},\mathbf{q}^{\prime}}=&&-\frac{1}{2}\,i\epsilon \, q\, q^{\prime}_{}\, e^{-i(\phi +\phi^{\prime}_{})}_{}\frac{\hbar}{2M\sqrt{\omega_{\mathbf{q}}^{(h)}\omega_{\mathbf{q}^{\prime}}^{(h)}}}\nonumber
\label{Elements}
\end{eqnarray} 
where $\epsilon =(3a/4)\partial t/\partial a$, $c^{\dagger}_{B,\mathbf{k}}$ denotes the creation operator of an electron in a Bloch state in sublattice $B$ and wave-vector $\mathbf{k}$, and $\phi$ ($\phi^{\prime}_{}$) is the angle of $\mathbf{q}$ ($\mathbf{q}^{\prime}_{}$) with respect to the $x$-axis.

These considerations give rise to an interesting competition: In-plane phonons are strongly coupled to electrons (the vertex is first order in the phonons) but they have a linear dispersion and hence a linearly vanishing density of states at small energy. In contrast, flexural phonons are weakly coupled to electrons (their vertex is second order in the phonons) but their dispersion is quadratic with constant density of states. In the following, we quantitatively analyze the consequences of this competition for the temperature dependence of the resistivity.
The latter is determined by 
\begin{equation}
\label{resistivity}
\rho =\frac{2}{e^{2}_{}v_{}^{2}\nu_{F}^{}}\,\frac{1}{\tau_{\mathrm{tr}}^{}}
\end{equation}
($\nu_{F}^{}$ is the electronic density of states at the Fermi level) in terms of the transport scattering rate
\begin{equation}
   \frac{1}{\tau_{\rm tr}}=\frac{2\pi}{\hbar} \sum_{f}^{} \left|M_{fi}^{}\right|^{2}_{}\left(1-\cos \theta\right)\delta(E_{f}-E_{i}\pm\hbar\omega)\; .
\label{Fermi}
\end{equation}
of Dirac fermions due to absorption (or emission) of phonons. Here, $\theta$ is the scattering angle of the Dirac fermions, $|i \rangle$ and $|f \rangle$ the initial and final scattering states, $E_{i}$ and $E_{f}$ the initial and final electronic energies, $\hbar\omega$ the energy of the absorbed (emitted) phonons, and $M_{fi}^{}=\langle f|H_{\rm ep}^{}|i \rangle$.

We consider the doped regime with $E_{F}^{}\gg k_{\mathrm{B}}^{}T$ and employ the conventional quasielastic approximation of neglecting the phononic contribution to energy conservation, as the typical phonon energies are small compared to $E_{F}^{}$. Moreover, we focus on sufficiently low temperatures so that we can restrict attention to the quadratic (linear) region of the dispersion of flexural (in-plane) phonons. 

{\em Flexural phonons}.---Scattering of Dirac fermions by flexural phonons requires absorption (or emission) of two phonons, say with wavenumbers $\mathbf{q}$ and $\mathbf{q}^{\prime}_{}$. Thus, the corresponding initial and final scattering states are $|i\rangle =|\mathbf{k},\sigma^{}_{}\rangle \otimes |n^{(h)}_{\mathbf{q}}\rangle$ and $|f\rangle =|\mathbf{k}^{}_{}+\mathbf{q}^{}_{}+\mathbf{q}^{\prime}_{},\sigma^{\prime}_{}\rangle \otimes |n^{(h)}_{\mathbf{q}}-1,n^{(h)}_{\mathbf{q}^{\prime}_{}}-1\rangle$, where $n^{(h)}_{\mathbf{q}}=\left[ \exp (\hbar\omega^{(h)}_{\mathbf{q}}/k_{\mathrm{B}}^{}T)-1\right]^{-1}_{}$ denotes the Bose distribution function of the flexural phonons and $|\mathbf{k}^{}_{},\sigma^{}_{}\rangle=1/\sqrt{2}\left(\sigma e^{-i\xi}_{}c^{\dagger}_{A,\mathbf{k}}+c^{\dagger}_{B,\mathbf{k}}\right)|\mathrm{vac}\rangle$ is a Dirac fermion state state with momentum $\mathbf{k}$ and chirality $\sigma$. (Here, $\xi$ is the angle between $\mathbf{k}$ and the $x$-axis and $|\mathrm{vac}\rangle$ is the electronic vacuum). Thus, $|M_{fi}^{}|^{2}_{}= |V^{(h)}_{\mathbf{q},\mathbf{q}_{}^{\prime}}|^{2}_{}\left[1- \sigma\sigma^{\prime}_{}\cos (\xi +\xi^{\prime}_{}+2\phi+2\phi^{\prime}_{})\right]/2$. The summation over final states requires integration over $\mathbf{q}$ and $\mathbf{q}_{}^{\prime}$, while averaging over the direction of the incoming electron will suppress the oscillatory term in $|M_{fi}^{}|^{2}_{}$. Using  $a^{(\nu )}_{\mathbf{q}}|n^{(\nu )}_{\mathbf{q}}\rangle =(n^{(\nu )}_{\mathbf{q}})^{1/2}_{}|n^{(\nu )}_{\mathbf{q}}-1\rangle$, $E_{i}^{}=\hbar vk=E_{F}^{}$, and $E_{f}^{}=\hbar v|\mathbf{k}+\mathbf{q}+\mathbf{q}^{\prime}_{}|$, we obtain 
\begin{equation}
\frac{1}{\tau_{\mathrm{tr}}^{(h)}}=\sum_{\mathbf{q},\mathbf{q}^{\prime}_{}}^{} \frac{\pi|V^{(h)}_{\mathbf{q},\mathbf{q}^{\prime}}|^{2}_{}}{\hbar}\left(1-\cos \theta\right)n^{(h)}_{q}n^{(h)}_{q^{\prime}_{}}\delta(\Phi_{\mathbf{q}+\mathbf{q}^{\prime}_{}}) \; ,
\label{Rateflex}
\end{equation}
where $\Phi_{\mathbf{q}+\mathbf{q}^{\prime}_{}}=\hbar v|\mathbf{k}+\mathbf{q}+\mathbf{q}^{\prime}_{}|-E_{F}^{}$ and $|V^{(h)}_{\mathbf{q},\mathbf{q}^{\prime}}|^{2}_{}=\hbar^{2}_{}\epsilon^{2}_{}/(16 M^{2}_{}\alpha^{2}_{})$ is independent of wavenumbers due to the quadratic dispersion of flexural phonons. The rate in Eq.\ (\ref{Rateflex}) is formally singular at small $q,q^{\prime}_{}$ since both Bose distributions diverge as $T/q^{2}_{}$. Rescaling momenta by $\sqrt{T}$ and introducing a cutoff $q_{c}$ at small wavenumbers (to be specified below), we obtain the scattering rate 
\begin{equation}
\label{RateflexT}
\frac{1}{\tau^{(h)}_{\rm tr}}\simeq\frac{C^{(h)}_{}}{32\, \pi^{2}}\frac{\hbar\omega^{(h)}_{k_{F}^{}}}{\kappa}\left(\frac{k_{\mathrm{B}}^{}T}{\hbar\omega_{k_{F}^{}}^{(h)}}\right)^{5/2}_{}\ln \left(\frac{k_{\mathrm{B}}^{}T}{\hbar \omega^{(h)}_{q_{c}^{}}}\right) \; ,
\end{equation}
where $C^{(\nu )}_{}=\epsilon^{2}_{}\hbar k^{4}_{F}/(4\rho_0 E_{F}^{}\hbar \omega^{(\nu )}_{k_{F}^{}})$. 
{\em The unusual $T^{5/2}_{}$ scaling implies that scattering from flexural phonons dominates the phonon contribution to the resistivity at low temperatures.} Indeed, the conventional phonon contribution to the resistivity due to in-plane phonons scales as $T^4$, which is the direct two-dimensional analog of the $T^5$ law in bulk solids.

Eq.\ (\ref{RateflexT}) shows that the scattering rate from flexural phonons diverges logarithmically for a strictly quadratic phonon dispersion. For clean elastic membranes, a low-momentum cutoff arises from the coupling terms between bending and stretching degrees of freedom of the membrane which are included in the Lagrangian Eq.\ (\ref{elastic}).
These lead to long-wavelength corrections to the elastic constants and hence the phonon dispersions \cite{Nelson}. Indeed, it is these renormalizations which are responsible for the stability of the flat phase at low temperatures. 

\begin{figure}[t]
	\centering
		\includegraphics[width=0.5\columnwidth]{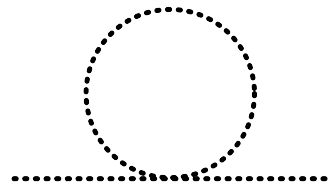}
			\caption{One-loop correction to the bending rigidity due to the effective interaction between flexural modes.
			\label{Loop}}
\end{figure}

In order to study the renormalization of the bending rigidity by the coupling of bending and stretching modes, we
integrate out the in-plane distortions in Eq.\ (\ref{elastic}) and obtain an effective energy functional for the flexural modes alone \cite{Nelson},
\begin{equation}
F=\frac{1}{2}\int \mathrm{d}\mathbf{r} \left[\kappa \left(\nabla^{2}_{}h\right)^{2}_{}+\frac{K_{0}^{}}{4}\left[ P^{\perp}_{\alpha\beta}\left(\partial_{\alpha}^{}h\right)\left(\partial_{\beta}^{}h\right)\right]^{2}_{}\right]\, ,
\label{Flexaction}
\end{equation}
where $K_{0}^{}=4\mu (\mu+\lambda)/(2\mu +\lambda)$ and $P^{\perp}_{\alpha\beta}=(\epsilon_{\alpha l}^{}\,\epsilon_{\beta k}^{}\, \partial_{l}^{}\, \partial_{k}^{})/\nabla^{2}_{}$ is a transverse projector. 
This effective energy functional contains a four-leg interaction between flexural distortions. Treating this 
quartic term to one-loop order (depicted in Fig.\ \ref{Loop}) in a conventional momentum-shell RG, one obtains the flow equation
\begin{equation}
\label{RG}
\frac{d\kappa}{dq}=-\frac{3}{16\pi}\,\frac{K_{0}^{}k_{\mathrm{B}}^{}T}{\kappa q^{3}_{}}\, ,
\end{equation}
with $q$ denoting the running shell wavevector. 

The flow equation Eq.\ (\ref{RG}) is readily solved, and the resulting scaling of $\kappa$ with $q$ is
\begin{equation}
\label{kappalambda}
\kappa(q)=\kappa\sqrt{1+q_{c}^{2}/q^{2}_{}}
\end{equation}
in terms of the temperature-dependent momentum scale $q_{c}^{}=[3K_{0}^{}k_{\mathrm{B}}^{}T/ (8\pi\kappa^{2}_{})]^{1/2}_{}$. Thus, the flexural phonon dispersion is quadratic for $q\gg q_c$. In contrast, for $q\ll q_c$, thermal fluctuations effectively stiffen the membrane and, within the simple one-loop analysis presented here, we find a renormalized dispersion $\omega_{q\ll q_{c}^{}}^{(h)}=(\kappa q_{c}^{})^{1/2}_{} q^{3/2}_{}$. It is this renormalization of the flexural-phonon dispersion at long wavelength which removes the singularity in the phonon scattering rate Eq.\ (\ref{RateflexT}). We therefore identify the low-momentum cutoff $q_c$ entering into Eq.\ (\ref{RateflexT}) with this momentum scale. 

As a result, we find that the temperature cancels from the argument of the logarithm in Eq.\ (\ref{RateflexT}). 
This analysis is approximate in that it is restricted to one-loop order and that it neglects the flow of the elastic constants for stretching deformations. While the stiffening against bending deformation would survive inclusion in a more refined treatment, the precise power-law dependence of $q_c$ on temperature would change. As a result, the argument of the logarithm would become temperature dependent, resulting in an overall $T^{5/2}\ln T$ scaling of the flexural phonon contribution to the resistivity.

{\em Crossover temperature.}---In order to estimate the crossover temperature $T_x$ below which phonon scattering of Dirac fermions is dominated by flexural modes, we note that an analogous calculation yields 
\begin{equation}
\label{RateinT}
\frac{1}{\tau_{\rm tr}^{(\nu )}}\simeq C^{(\nu )}_{}\,\left(\frac{k_{\mathrm{B}}^{}T}{\hbar\omega^{(\nu )}_{k_{F}^{}}}\right)^{4}
\end{equation} 
for the transport scattering rate from in-plane phonons ($\nu = l,t$). Thus, the crossover temperature obtained from a comparison of Eqs.\ (\ref{RateflexT}) and (\ref{RateinT}) becomes
\begin{equation}
\label{Tx}
  T_{x}^{} = \frac{1}{k_{\mathrm{B}}}\left(\frac{\ln  \left( k_{\mathrm{B}}^{}T/\hbar \omega^{(h)}_{q_{c}^{}}\right)}{8\, (2\pi )^{2}_{}}\, \frac{\hbar\omega^{(t)\, 5}_{k_{F}^{}}}{\kappa \left(\hbar\omega^{(h)}_{k_{F}^{}}\right)^{5/2}}\right)^{2/3}_{}.
\end{equation}
This result is independent of the Fermi energy (i.e.\ the doping level) and with typical parameters for graphene \cite{Parameters}, we obtain $T_{x}^{}\simeq 70\, \mathrm{K}$. There is therefore a significant temperature range accessible in experiment over which our predictions for the resistivity can be tested.

{\em Conclusions}.---For clean graphene membranes, long-wavelength renormalizations of their elastic properties due to thermal fluctuations are crucial in order to obtain a finite transport scattering rate and hence resistivity $\rho\propto T^{5/2}_{}\ln T$ from scattering by flexural phonons. In the presence of disorder, the elastic properties of the membrane are renormalized even at zero temperature \cite{Radzihovsky,Lubensky}. Indeed, the recent experimental observations of rippling \cite{GeimRipples} suggest that disorder, inside or close to the 2d membrane, exists in present graphene membranes. Such disorder-induced renormalizations of the elastic moduli may compete with the renormalizations by thermal fluctuations and lead to a temperature-independent saturation of the cutoff $q_c$ at low temperatures. A detailed study of the effects of disorder in this context remains an important topic for future research. 

{\em Acknowledgments}---We are grateful to F.\ Guinea, T.\ Nattermann, L.\ Peliti, and Ady Stern for instructive discussions. This work was supported in part by DIP. FvO enjoyed the hospitality of the KITP Santa Barbara (supported in part by NSF grant PHY99-07949) and EM acknowledges the hospitality of the Weizmann Institute of Science (supported by grant RITA-CT-2003-506095) .


\begin{thebibliography}{99}
\bibitem{Parameters} Typical parameters for graphene sheets are: $\kappa =1\,\mathrm{eV}$,  $\mu \simeq 4\lambda =9\,\mathrm{eV}\AA^{-2}_{}$, $\hbar \alpha =35\,\mathrm{meV}\AA^{2}_{}$.
\end{thebibliography}
\end{document}